\documentclass[11pt]{article}

\usepackage[colorlinks,bookmarksopen,bookmarksnumbered,citecolor=red,urlcolor=red]{hyperref}

\usepackage{natbib}
\usepackage{amsmath}
\usepackage{amssymb}
\usepackage{amsfonts}
\usepackage{amsthm}
\usepackage{graphicx}
\usepackage[algo2e,ruled]{algorithm2e}
\usepackage{dsfont}
\usepackage{mathrsfs}
\usepackage{array}
\usepackage{colortbl}
\usepackage[format=hang, font=small, labelfont=bf]{caption}
\usepackage{pdfpages}
\usepackage{float}

\usepackage{fullpage}
\usepackage{authblk}

\bibpunct{(}{)}{,}{a}{,}{,}

\newcommand{\eg}{\textit{e.g.}}
\newcommand{\ie}{\textit{i.e.}}

\newcommand{\suchthat}{\;\ifnum\currentgrouptype=16 \middle\fi|\;}

\newcommand{\s}{\mathscr{S}}
\newcommand{\G}{\mathscr{G}}

\newcommand{\Xv}{\mathbf{X}}
\newcommand{\x}{\mathbf{x}}
\newcommand{\X}{\mathscr{X}^n}
\newcommand{\Y}{\mathscr{Y}}
\newcommand{\y}{\mathbf{y}}
\newcommand{\Yv}{\mathbf{Y}}
\newcommand{\yobs}{\mathbf{y}^{\text{obs}}}

\newcommand{\ivj}{i\stackrel{\G}{\sim} j}

\newcommand{\N}{\mathscr{N}}

\newcommand{\mle}{\text{MLE}}

\newcommand{\BIC}{\text{BIC}}
\newcommand{\PLIC}{\text{PLIC}}
\newcommand{\BLIC}{\text{BLIC}}

\newcommand{\mfl}{\text{MF-like}}

\newcommand{\gbf}{\text{GBF}}

\newcommand{\pr}{\mathbf{P}}

\DeclareMathOperator*{\gO}{\mathcal{O}}

\newcommand{\vertiii}[1]{{\left\vert\kern-0.25ex\left\vert\kern-0.25ex\left\vert #1 
    \right\vert\kern-0.25ex\right\vert\kern-0.25ex\right\vert}}

\newcommand{\KL}{\text{KL}}

\newcommand{\Ss}{\mathbf{S}}

\newcommand{\indxixj}{\mathds{1}\{x_i=x_j\}}

\begin{document}

\title{Hidden Gibbs random fields model selection using Block Likelihood Information Criterion}

\author[1]{Julien Stoehr}

\author[2]{Jean-Michel Marin\footnote{jean-michel.marin@umontpellier.fr}}

\author[3]{Pierre Pudlo}

\affil[1]{School of Mathematical Sciences \& Insight Centre for Data Analytics, University College Dublin, Ireland}
\affil[2]{Institut Montpelli\'erain Alexander Grothendieck, Universit\'e de Montpellier, France}
\affil[3]{Institut de Math\'ematiques de Marseille, Aix-Marseille Universit\'e, France}

\maketitle

\begin{abstract}
Performing model selection between Gibbs random fields is a very challenging task. Indeed, due to the Markovian dependence structure, the normalizing constant of the fields cannot be computed using standard analytical or numerical methods. Furthermore, such unobserved fields cannot be integrated out and the likelihood evaluztion is a doubly intractable problem. This forms a central issue to pick the model that best fits an observed data.
We introduce a new approximate version of the \textit{Bayesian Information Criterion}. We partition the lattice into continuous rectangular blocks and we approximate the probability measure of the hidden Gibbs field by the product of some Gibbs distributions over the blocks. On that basis, we estimate the likelihood and derive the \textit{Block Likelihood Information Criterion} (BLIC) that answers model choice questions such as the selection of the dependency structure or the number of latent states. We study the performances of BLIC for those questions. In addition, we present a comparison with ABC algorithms to point out that the novel criterion offers a better trade-off between time efficiency and reliable results.

\vspace{0.5cm} \noindent \textbf{Keywords:} Hidden Markov random fields; model selection; Bayesian Information Criterion
\end{abstract}

\section{Introduction}

Gibbs or discrete Markov random fields have appeared as convenient statistical model to analyse different types of spatially correlated data. Notable examples are the autologistic model \citep{besag74} and its extension the Potts model used to describe the spatial dependency of discrete random variables (\textit{e.g.}, shades of grey or colors) on the vertices of an undirected graph (\textit{e.g.}, a
regular grid of pixels).  In particular, hidden Markov random fields offer an appropriate representation for practical settings where the true state is unknown. The general framework can be described as an observed data $\y$ which is a noisy or incomplete version of an unobserved discrete latent process $\x$. Shaped by the development of \cite{geman1984} and \cite{besag1986}, these models have enjoyed great success in image analysis -- see for example \citet{alfo08} and \citet{moores14} who performed image segmentation with the help of this modelling --  but also in other applications including disease mapping \citep[\textit{e.g., }][]{green02} and genetic analysis \citep[\textit{e.g., }][]{francois06, friel2009} to name a few. Despite their popularity,  Gibbs random fields suffer from major computational difficulties since their normalizing constant is intractable. This forms a central issue in statistical analysis as the computation of the likelihood is an integral part of the procedure for both parameter inference \citep[\textit{e.g., }][]{celeux2003, friel2009, mcgrory2009, everitt12} and model selection \citep[\textit{e.g., }][]{grelaud09, frielevidence, cucala13, stoehr2015}. Remark the exception of small latices on which
we can apply the recursive algorithm of \cite{reeves2004, friel2007} and obtain an exact computation of the normalizing constant. However, the complexity in time of the above algorithm grows exponentially and is thus helpless on large lattices.

The present paper cares about the problem of selecting the number of latent states as well as the dependency structure of hidden Potts model and explores the opportunity of using the Bayesian Information Criterion \citep[BIC,][]{schwarz1978} to answer the question. If the problem of recovering the number of hidden states is common in image segmentation, the problem of selecting a dependency structure has received little attention in the literature. \cite{stoehr2015} have proposed to use approximate Bayesian computation (ABC) model choice \citep[\textit{e.g., }][]{surveyABC} based on geometric summary statistics to tackle the choice of an underlying graph but their approach is restricted to the latter. While our work is motivated by a more general issue, it offers a way to overcome the computational burden of ABC algorithms. 

Model choice is a problem of probabilistic model comparison. 
The standard approach to compare one model against another is based on the Bayes factor \citep{kass1995} that involves the ratio of the evidence of each model. However the evidence can usually not be computed with standard procedure due to a high-dimensional integral. Various approximations have been proposed but a commonly used one, if only for its simplicity, is BIC that is an asymptotic estimate of the evidence based on the Laplace method for integrals. The criterion is a simple penalized function of the maximized log-likelihood which, in the context of hidden Gibbs random fields, cannot be computed since it requires to integrate the intractable Gibbs distribution over the latent space configurations. As regards the simpler case of observed Markov random field solutions have been brought by penalized pseudolikelihood \citep{ji1996} or MCMC approximation of BIC \citep{seymour1996}. To circumvent the computational difficulties in the hidden case, little has been done before the work of \cite{stanford2002} and \cite{forbes2003}. Both propose approximations that consist in replacing the intractable likelihood with a product distribution on system of independent variables to make the computation tractable.
Our main contribution is to show that larger collections of variables, namely blocks of the lattice, can be considered by taking advantage of the exact recursion of \cite{reeves2004} and leads to an efficient criterion : the Block Likelihood Information Criterion (BLIC). In particular, we will show that a reasonable approximation of the Gibbs distribution is a product of Gibbs distributions on each independent block. Such ideas have occurred in the context of composite likelihood but the use of non-genuine probability distribution results in misspecified model \citep[\textit{e.g.,}][]{okabayashi2011, frielproc, stoehrfriel2015} that we have decided to avoid. 

The paper is organized as follows: Section \ref{sec:hmrf} presents hidden Gibbs random fields. In section \ref{sec:BLIC}, after recalling the basis of BIC, we introduced our Block Likelihood Information Criterion (BLIC). In Section \ref{sec:blic-results}, to assess the performances of the novel criterion, it is compared to pre-existing criteria on simulated data sets. We fill in our study with a comparison between BLIC and the ABC algorithm of \cite{stoehr2015}.

\section{Hidden Gibbs random fields}
\label{sec:hmrf}

A discrete random field $\Xv$ is a collection of random variables $X_i$ indexed by a finite set  $\s = \{1,\dots,n\}$, whose elements are called sites, and taking values in a finite state space $\mathscr{X}:=\{0,\ldots,K-1\}$, interpreted as colors. For a given subset $A\subset\s$, $\Xv_A$ and $\x_A$ respectively define the random process on $A$, \ie, $\{X_i, i\in A\}$, and a realisation of $\Xv_A$. Denotes $\s\setminus A= -A$ the complement of $A$ in $\s$.
When modeling a digital image, the sites are lying on a
regular 2D-grid of pixels, and their dependency is given by an
undirected graph $\G$ which induces a topology on $\s$: by definition, sites $i$ and $j$ are adjacent or neighbor if and only if $i$ and $j$ are linked by an edge in $\G$. A random field $\Xv$ is a Markov random field with respect to $\G$, if for all configuration $\x$ and for all sites $i$
\begin{equation}
\pr\left(X_i = x_i\suchthat \Xv_{-i} = \x_{-i}\right) = \pr\left(X_i = x_i\suchthat \Xv_{\N(i)} = \x_{\N(i)}\right),
\label{eqn:markov}
\end{equation}
where $\N(i)$ denotes the set of all the adjacent sites to $i$ in $\G$.
The Hammersley-Clifford theorem states that if the distribution of a Markov random field with respect to a graph $\G$ is positive for all configuration $\x$ then it admits a Gibbs representation for the same topology (see for example \cite{grimmett1973, besag74} and for a historical perspective \cite{clifford1990}),  namely a probability measure $\pi$ on $\X$ given by
\begin{equation}
\pi\left(\x\suchthat \psi, \G\right) = \frac{1}{Z\left(\psi, \G\right)}\exp\left\lbrace- H\left(\x\suchthat\psi,\G\right)\right\rbrace,
\label{eqn:gibbs}
\end{equation}
where $\psi = (\psi_1,\ldots, \psi_d)$ is a vector of parameters, $H$ denotes the energy function or Hamiltonian. The present paper solely focuses on models whose Hamiltonian linearly depends on the parameter $\psi$, that is
\[
H\left(\x\suchthat\psi,\G\right) = -\psi^{T}\Ss(\x).
\]
where $\Ss(\x)=(s_1(\x),\ldots,s_d(\x))$ is a vector of sufficient statistics. The inherent difficulty of all these models that arises from the intractable normalizing constant, called the partition function, defined by
\[
Z(\psi, \G)  = \sum_{\x\in \X} \exp\left\lbrace \psi^{T}\Ss(\x)\right\rbrace
\]
The latter is a summation over the numerous possible realizations of the random field $\Xv$, that cannot be computed directly (except for small grids and small number of colors $K$). 

In hidden Markov random fields, the latent process is observed indirectly through another field; this permits the modelling of noise that may happen upon many concrete situations. The aim is to infer some properties of a latent state $\x$ given an observation $\y$. Precisely, given the realization $\x$ of the latent, the observation $\y$ is a family of random variables indexed by the set of sites $\s$, and taking values in a set $\Y$, \textit{i.e.}, $\y=\left\lbrace y_i; {i\in\s}\right\rbrace$, and are commonly assumed as independent draws that form a noisy version of the hidden field. Consequently, we set the conditional distribution of
$\Yv$ knowing $\Xv=\x$, also called emission distribution, as the product 
\[
\pi\left(\y\suchthat \x, \phi\right)=\prod_{i\in\s}
\pi\left(y_i\suchthat x_i,\phi\right), 
\]
where $\pi(y_i\mid x_i,\phi)$ is the marginal noise distribution parametrized by $\phi$, that is given for any site $i$. Those marginal distributions are for instance discrete distributions \citep{everitt12}, Gaussian \citep[\textit{e.g., }][]{besag1991, qian1991, forbes2003, cucala13} or Poisson distributions \citep[\textit{e.g., }][]{besag1991}. Model of noise that takes into account information of the nearest neighbors have also been explored \citep{besag1986}. Hence the likelihood of the
hidden Gibbs random field with parameter $\psi$ on the graph $\G$ and emission 
distribution $\pi\left(\cdot\suchthat \x, \phi\right)$ is given by 
\begin{equation}
\pi\left(\y\suchthat \phi, \psi\right) = \sum_{\X} \pi\left(\y\suchthat \x, \phi\right)\pi\left(\x\suchthat \psi, \G\right).
\label{eqn:int-latent}
\end{equation}
The latter faces a double intractable issue as neither the likelihood of the latent field, nor the above sum can be computed directly: the
cardinality of the range of the sum is of combinatorial complexity. 

\section{Block Likelihood Information Criterion}
\label{sec:BLIC}

The Bayesian Information Criterion offers a mean arising from Bayesian viewpoint to select a statistical model. In what follows, we provide solely the foundation that motivates our contribution and we refer the reader for instance to \cite{raftery1995bic} for a more detailed presentation. 

\subsection{Background on Bayesian Information Criterion}

We are given $n$ independent and identically distributed observations $\y=\left\lbrace y_1,\ldots,y_n\right\rbrace$ from an unknown statistical model to estimate. The Bayesian approach to model selection is based on posterior model probabilities. Consider a finite set of models $\left\lbrace m : 1,\ldots,M\right\rbrace$ where each one is defined by a probability density function $\pi_m$ related to a parameter space $\Theta_m$. The model that best fits an observation $\y$ is the model with the highest posterior probability
\[
\pi\left(m\suchthat \y\right) = \frac{\pi(m)e\left(\y\suchthat m\right)}{\sum_{m'}\pi(m')e\left(\y\suchthat m'\right)},
\]
where $e\left(\y\suchthat m\right)$ denotes the evidence of $m$, that is the joint distribution of $(\y,\theta_m)$ integrated over space parameter $\Theta_m$
\begin{equation*}
e\left(\y\suchthat m\right) = \int \pi_m\left(\y\suchthat\theta_m\right)\pi_m\left(\theta_m\right)\mathrm{d}\theta_m.
\end{equation*}
Under the assumption of model being equally likely \textit{a priori}, it is equivalent to choose the model with the largest evidence. From the Laplace method for integrals, under regularity conditions, the evidence of model $m$ can be written as
\begin{equation}
\log e\left(\y\suchthat m\right) = \log \pi_m\left(\y\suchthat \hat{\theta}_{\mle}\right) - d_m\log(n) + R_m\left(\hat{\theta}_{\mle}\right) + \gO\left(n^{-\frac{1}{2}}\right),
\label{eqn:laplace-evidence}
\end{equation}
where where $\hat{\theta}_{\mle}$ is the maximum likelihood estimator of $\pi_m$, $d_m$ is the number of free parameters for model $m$ and $R_m$ is bounded as the sample size grows to infinity \citep[\eg, ][]{schwarz1978, tierney1986}.

BIC is an asymptotical estimate of the evidence defined by
\begin{equation}
-2\log e\left(\y\suchthat m\right) \simeq \BIC(m) = -2\log \pi_m\left(\y\suchthat \hat{\theta}_{\mle}\right) + d_m\log(n).
\label{eqn:bic-iid}
\end{equation}
The $d_m\log(n)$ term corresponds to a penalty term which increases with the complexity of the model. Thus selecting the model with the largest evidence is equivalent to choose the model which minimizes BIC. Regardless of the prior on parameter, the error in \eqref{eqn:bic-iid} is, in general, solely bounded and does not go to zero even with an infinite amount of data. The approximation may hence seem somewhat crude. However as observed by \cite{kass1995} the criterion does not appear to be qualitatively misleading as long as the sample size $n$ is much larger than the number $d_m$ of free parameters in the model. In addition, a reasonable choice of the prior can lead to much smaller error. Indeed, \cite{kass1995reference} have found that the error is $\gO\left(n^{-1/2}\right)$ for a well chosen multivariate normal prior distribution.

BIC can be defined beside the special case of independent random variables. In the latter case the number of free parameter is, in general, not equal to the dimension of the parameter space as for the independent case. The consistency of BIC has been proven in various situations such as independent and identically distributed processes from the exponential families \citep{haughton1988}, mixture models \citep{keribin2000}, Markov chains \citep{csiszar2000, gassiat2002}. When dealing with observed Markov random fields, aside from the problem of intractable likelihoods the number of free parameters in the penalty term has no simple formula. In the context of selecting a neighborhood system, \cite{csiszar2006} proposed to replace the likelihood by the pseudolikelihood \citep{besag75} and modify the penalty term as the number of all possible configurations for the neighboring sites. The resulting criterion is shown to be consistent as regards this model choice. Up to our knowledge such a result has not been yet derived for hidden Markov random field. The problem of approximating  BIC could be termed a triple intractable problem since neither the maximum likelihood estimate $\hat{\theta}_{\mle}$ nor the incomplete likelihood $\pi_m(\cdot\mid \theta)$ can be computed with standard methods since they require to integrate over the latent configuration space and no simple definition of $d_m$ is available.

\subsection{Gibbs distribution approximations}
\label{subsec:gibbs-approx}

A convenient way to circumvent the issues of computing BIC is to replace the Gibbs distribution by tractable surrogates since it avoids the use of time consuming simulations methods. As for the pseudolikelihood \citep{besag75} and more generally composite likelihood \citep{lindsay1988}, the main idea consists in replacing the original Markov distribution by a product of easily normalized distribution. But while composite likelihoods are not a genuine probability distribution for Gibbs random field, the focus hereafter is solely on valid probability function by considering system of independent variables. This choice is motivated by the observations that at finite sample size, when dealing with composite likelihood, misspecification of the model has to be taken into account \citep[\textit{e.g.,}][]{frielproc, stoehrfriel2015}, so that constant terms may appear in the remainder $R_m$ in \eqref{eqn:laplace-evidence}.

Finding good approximations of the Gibbs distribution has long standing antecedents in statistical mechanics when one aims at predicting the response to the system to a change in the Hamiltonian. One important technique is based on a variational approach as the minimizer of the free energy, sometimes referred to as variational or Gibbs free energy and defined with the Kullback-Leibler divergence between $\pr$ and the target distribution $\pi(\cdot\mid\psi,\G)$ as
\begin{equation}
F(\pr)=-\log Z\left(\psi, \G\right) + \KL\left(\pr,\pi(\cdot\mid\psi,\G)\right).
\label{eqn:free-energy}
\end{equation}
The Kullback-Leibler divergence being non-negative and zero if and only if $\pr = \pi(\cdot\mid\psi,\G)$, the free energy has an optimal lower bound achieved for $\pr = \pi(\cdot\mid\psi,\G)$. Minimizing the free energy with respect to the set of probability distribution on $\X$ allows to recover the Gibbs distribution but presents the same computational intractability. A solution is to minimize the Kullback-Leibler divergence over a restricted class of tractable probability distribution on $\X$. This is the basis of mean field approaches that aim at minimizing the Kullback-Leibler divergence over the set of probability functions that factorize on sites of the lattice. The minimization of \eqref{eqn:free-energy} over this set leads to fixed point equations for each marginal of $\pr$ \citep[see for example][]{jordan1999}. The resulting solution motivates the mean field-like approximations of \cite{celeux2003} for which the neighbors of a site $i$ are set to well chosen constant independently of the value at the given site, namely
\begin{equation}
\pr^{\mfl}\left(\x\suchthat\psi,\G\right)= \prod_{i\in\s}\pi\left(x_i\suchthat\Xv_{\N(i)}=\tilde{\x}_{\N(i)},\psi,\G\right).
\end{equation}

Instead of considering distributions that completely factorize on single sites, we are hereafter interested in tractable approximations that factorize over larger sets of nodes, namely blocks of the lattice. Consider a partition of $\s$ into contiguous rectangular blocks, namely
\[
\s = \bigsqcup_{\ell=1}^C A(\ell),
\] 
and denote $\tilde{D}$ the class of independent probability distributions $\pr$ that factorize with respect to this partition, that is if $\X_{A(\ell)} $ stands for the configuration space of the block $A(\ell)$, for all $\x$ in $\X$
\[
\pr(\x)=\prod_{\ell = 1}^C \pr_\ell\left(x_{A(\ell)}\right), \text{ where } \pr_{\ell}\in\mathcal{M}_{1}^{+}\left(\X_{A(\ell)}\right)
\text{ and }\pr\in\mathcal{M}_1^{+}(\X). 
\]
To take over from the Gibbs likelihood, we propose to explore the opportunity of probability distributions in $\tilde{D}$ of the form
\begin{equation}
\pr\left(\x\suchthat\tilde{\x}, A(1),\ldots,A(C),\psi\right) = \prod_{\ell = 1}^C \pi\left(\x_{A(\ell)}\suchthat \Xv_{B(\ell)} = \tilde{\x}_{B(\ell)}, \psi,\G\right),
\label{eqn:product-distrib}
\end{equation}
where $\tilde{\x}$ is a constant field in $\X$ to specify and $B(\ell)$ is either the border of $A(\ell)$, \ie, elements of the absolute complement of $A(\ell)$ that are connected to elements of $A(\ell)$ in $\G$, or the empty set. In the latter case, we are cancelling the edges in $\G$ that link elements of $A(\ell)$ to elements of any other subset of $\s$ such that the factorization is independent of $\tilde{\x}$. The Gibbs distribution is then simply replaced by the product of the likelihood restricted to $A(\ell)$. For instance a Potts model on $\X$ is replaced with a product of Potts models on $\X_{A(\ell)}$. To underline that point, $\tilde{\x}$ is omitted in what follows when $B(\ell)=\emptyset$. Note that composite likelihoods differs from \eqref{eqn:product-distrib} in most cases since blocks are not allowed to overlap and contrary to conditional composite likelihoods, neighbors are set to constants. The only example of composite likelihoods that lies in $\tilde{D}$ is marginal composite likelihoods for non overlapping blocks.

The assumption of independent blocks leads to tractable BIC approximations. Indeed, plugging the probability distribution \eqref{eqn:product-distrib} in place of the Gibbs distribution in \eqref{eqn:int-latent} yields
\begin{align}
\pr_m\left(\y\suchthat \tilde{\x}, \theta\right) & = \sum_{\x\in\X}\pi\left(\y\suchthat \x, \phi\right)\pr\left(\x\suchthat\tilde{\x}, A(1),\ldots,A(C),\psi\right) \nonumber \\
& = \prod_{\ell=1}^C~\sum_{\x_{A(\ell)}} \left\lbrace \prod_{i\in A(\ell)}
\pi\left(y_i\suchthat x_i, \phi\right)\right\rbrace 
\pi\left(\x_{A(\ell)}\suchthat \Xv_{B(\ell)} = \tilde{\x}_{B(\ell)}, \psi,\G\right) \nonumber \\
&=  \prod_{\ell=1}^C\sum_{\x_{A(\ell)}} 
\pi\left(\y_{A(\ell)}\suchthat \x_{A(\ell)}, \phi\right) \pi\left(\x_{A(\ell)}\suchthat \Xv_{B(\ell)} = \tilde{\x}_{B(\ell)}, \psi,\G\right).
\label{eqn:approx-incomplete-like}
\end{align}
This estimate of the incomplete likelihood $\pi_m(\cdot\mid\theta)$ leads to the following BIC approximations
\begin{equation}
\BIC(m) \approx -2\log\pr_m\left(\y\suchthat \tilde{\x}, \theta^{\ast}\right)+ d_m\log(\vert\s\vert) := \BLIC^{~\tilde{\x}}\left(m\suchthat\theta^{\ast}\right),
\label{eqn:BIC-gene}
\end{equation}
where $\theta^{\ast}=\left(\phi^{\ast},\psi^{\ast}\right)$ is a parameter value to specify. We refer to these approximations as Block Likelihood Information Criterion (BLIC). In the first instance, the number of free parameters $d_m$ is set to the dimension of $\Theta_m$, that is we are neglecting the interaction between variables within a block in the penalty term.

Our proposal relies on that each term of the product \eqref{eqn:approx-incomplete-like} can be computed using the recursion of \cite{friel2007} as long as the blocks are small enough. Indeed for models whose potential linearly depends on the parameter, the probability distribution on $A(\ell)$ can be written as a Gibbs distribution on the block conditioned on the fixed border $\tilde{\x}_{B(\ell)}$, namely
\[
\pi\left(\x_{A(\ell)}\suchthat \Xv_{B(\ell)} = \tilde{\x}_{B(\ell)}, \psi,\G\right) = 
\frac{1}{Z\left(\psi,\G, \tilde{\x}_{B(\ell)}\right)}\exp\left\lbrace\psi^T\Ss\left(\x_{A(\ell)}\suchthat\tilde{\x}\right)\right\rbrace,
\]
where $\Ss\left(\x_{A(\ell)}\suchthat\tilde{x}\right)$ is the restriction of $\Ss$ to the subgraph defined on the set $A(\ell)$ and conditioned on the fixed border $\tilde{\x}_{B(\ell)}$, and $Z\left(\psi,\G, \tilde{\x}_{B(\ell)}\right)$ is the corresponding normalizing constant. Assuming that all the marginals of the emission distribution are positive, it follows
\begin{multline*}
\sum_{\x_{A(\ell)}} 
\pi\left(\y_{A(\ell)}\suchthat \x_{A(\ell)}, \phi\right) \pi\left(\x_{A(\ell)}\suchthat \Xv_{B(\ell)} = \tilde{\x}_{B(\ell)}, \psi,\G\right) 
\\ =
\frac{1}{Z\left(\psi,\G, \tilde{\x}_{B(\ell)}\right)}\underbrace{\sum_{\x_{A(\ell)}} \exp\left\lbrace 
\log \pi\left(\y_{A(\ell)}\suchthat \x_{A(\ell)}, \phi\right) + \psi^T\Ss\left(\x_{A(\ell)}\suchthat\tilde{\x}\right)
\right\rbrace}_{=Z\left(\theta,\G,\y_{A(\ell)},\tilde{\x}_{B(\ell)}\right)}.
\end{multline*}
The term $Z\left(\theta,\G,\y_{A(\ell)},\tilde{\x}_{B(\ell)}\right)$ corresponds to the normalizing constant of the conditional random field $\Xv_{A(\ell)}$ knowing $\Yv_{A(\ell)}=\y_{A(\ell)}$ and $\Xv_{B(\ell)}=\tilde{\x}_{B(\ell)}$, that is the initial model with an extra potential on singletons. 
Then the algebraic simplification at the core of the algorithm of \cite{friel2007} applies for both normalizing constants, such that we can exactly compute the Block Likelihood Information Criterion, namely
\begin{equation}
\BLIC^{~\tilde{\x}}\left(m\suchthat\theta^{\ast}\right) = -2\sum_{\ell=1}^C \bigg\lbrace\log Z\left(\theta^{\ast},\G,\y_{A(\ell)},\tilde{\x}_{B(\ell)}\right)
-\log Z\left(\psi^{\ast},\G, \tilde{\x}_{B(\ell)}\right)\bigg\rbrace + d_m\log(\vert\s\vert).
\label{eqn:bclic}
\end{equation}

\subsection{Related model choice criteria}

This approach encompasses the Pseudolikelihood Information Criterion \\ \citep[PLIC,][]{stanford2002} as well as the mean field-like approximations $\BIC^{\mfl}$ proposed by \cite{forbes2003}. When one considers the finest partition of $\s$, that is distributions that factorize on sites, they have already proposed ingenious solutions for choosing $\tilde{\x}$ and estimating $\hat{\theta}_{\ast}$ in \eqref{eqn:BIC-gene}. Indeed, \cite{stanford2002} suggest to set $(\tilde{\x}, \hat{\theta}_{\ast})$ to the final estimates $(\hat{\theta}^{\text{ICM}},\tilde{\x}^{\text{ICM}})$ of the unsupervised Iterated Conditional Modes \citep[ICM,][]{besag1986} algorithm, while \cite{forbes2003}  put forward the use of the output $(\hat{\theta}^{\mfl},\tilde{\x}^{\mfl})$ of the simulated field algorithm  of \cite{celeux2003}. To make this statement clear, we could note
\begin{align*}
\PLIC(m) & = \BLIC^{~\tilde{\x}^{\rm ICM}}\left(m\suchthat \hat{\theta}^{\rm ICM}\right), \\ 
\BIC^{\mfl}(m) & = \BLIC^{~\tilde{\x}^{\mfl}}\left(m\suchthat \hat{\theta}^{\mfl}\right).
\end{align*}
Whilst PLIC shows good result as regards the selection of the number of components of the hidden state, ICM performs poorly for the parameter estimation in comparison with the EM-like algorithm of \cite{celeux2003}. Hence we advocate in favour of the latter in what follows to get estimates of $\hat{\theta}_{\mle}$ and to fix a segmented random field $\tilde{\x}$. 

We shall also remark that for a factorization over the graph nodes when $B(\ell)=\emptyset$ we retrieve a mixture model. Indeed, turning off all the edges in $\G$ leads to approximate the Gibbs distribution by a multinomial distribution with event probabilities depending on the potential on singletons. Hence if marginal emission distribution are Gaussian random variables depending on the component on the latent site associated, we would deal with a classical Gaussian mixture model.

\section{Comparison of BIC approximations}
\label{sec:blic-results}

Our primary intent with the BIC approximations was to choose the number of latent states as well as the dependency structure of a hidden Markov random fields. The following numerical experiments illustrate the performances as regards these questions for realizations of a hidden Potts model.

\subsection{Hidden Potts models}
\label{sub:bic-Potts}

This numerical part of the paper focuses on observations for which the hidden field is modelled by a $K$-states Potts model. While being widely used in practice \citep[\eg,][]{hurn03,alfo08,francois06,moores14}, the model is representative of the computational difficulties of hidden Gibbs random field. The model sets a probability distribution on $\X=\left\lbrace1,\ldots,K\right\rbrace^n$ parametrized by a scalar $\psi$ that adjusts the level of dependency between adjacent sites and whose Hamiltonian is given by 
\[
H\left(\x\suchthat\psi,\G\right) = -\psi\sum_{\ivj}\indxixj.
\]
The above sum $\ivj$ ranges the set of edges of the graph $\G$. In the statistical physic literature, $\psi$ is interpreted as the inverse of a temperature, and when the temperature drops below a fixed threshold, values $x_i$ of a typical realization of the field are almost all equal (the model then exhibits strong dependency between all sites). These peculiarities of Potts models are called phase transitions.
\begin{figure}[H]
\centering
\begin{minipage}[t]{7cm}
\centering
\includegraphics [height = 3cm, width = 3cm]{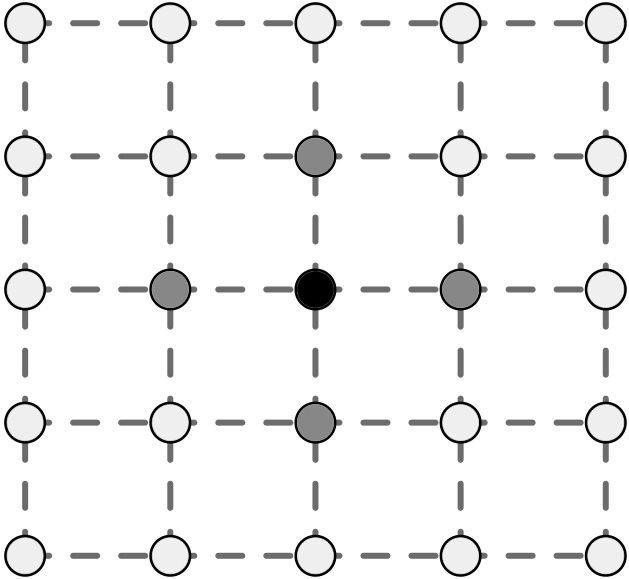}\\
(a)
\end{minipage}
\begin{minipage}[t]{7cm}
\centering
\includegraphics [height = 3cm, width = 3cm]{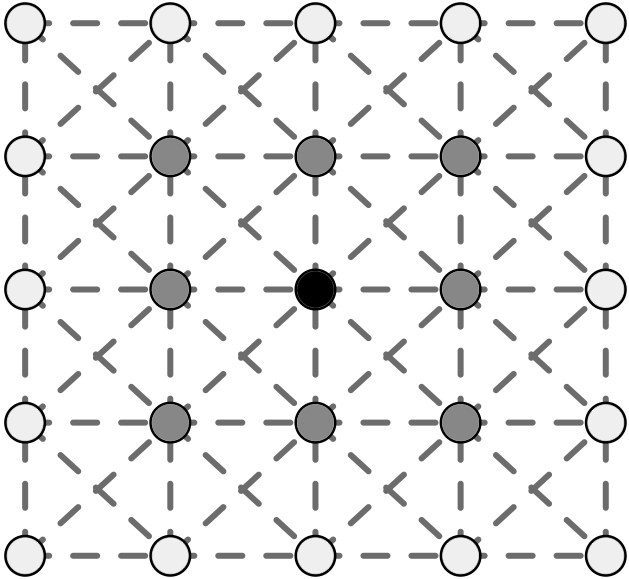}\\
(b)
\end{minipage}
\caption{Neighborhood graphs $\G$ of hidden Potts model. (a) The four closest neighbour graph $\G_4$ defining model $\text{HPM}(\G_4,\theta,K)$. (b) The eight closest neighbour graph $\G_8$ defining model $\text{HPM}(\G_8,\theta,K)$.}
\label{fig:neigh}
\end{figure}

We set the emission distribution
such that the marginal distribution are Gaussian distribution entered at the value of the related nodes, namely
\[
y_i\mid x_i = k~\sim \mathcal{N}\left(k,\sigma_k^2\right)\quad k\in \{0,\ldots,K-1\},
\]
where $\sigma_k$ is the standard deviation for sites belonging to class $k$. Even though the noise model is homoscedastic, we still index the standard deviation by $k$ since we do not use the assumption of a constant variance in the estimation procedure, such that the number of parameters estimated is $d_m = 2\times k +1$. The parameter to be estimated with the ICM or simulated field algorithms is then 
$
\theta=\left(\phi,\psi\right),~\text{with}~\phi=\left\lbrace
\left(k,\sigma_k\right):~k=0,\ldots,K-1
\right\rbrace.
$
We denote ${\rm HPM}(\G,\theta, K)$, the hidden K-states Potts model defined above.

The common point of our examples is to select the hidden Potts model that better fits a given observation $\yobs$ composed of $n=100\times 100$ pixels among a collection
\[
\mathcal{M}=\left\lbrace {\rm HPM}\left(\G,\theta, K\right):~K=K_{\rm min},\ldots,K_{\rm max}~;~\G\in\left\lbrace \G_4,\G_8\right\rbrace \right\rbrace,
\]
where $K$ is the number of colors of the corresponding model and $\G$ is one of the two possible neighborhood systems: $\G_4$ and $\G_8$, see Figure \ref{fig:neigh}. 
For each model ${\rm HPM}\left(\G,\theta, K\right)$, the estimate $\hat{\theta}_{\mle}$ and the segmented field $\tilde{\x}$ were computed using $\rm SpaCEM^3$ (see the Documentation on \url{http://spacem3.gforge.inria.fr}). The software allows the implementation of the unsupervised ICM algorithm as well as the simulated field algorithm and provides computation of PLIC, the mean field-like approximations $\BIC^{\mfl}$ and $\BIC^{\gbf}$. The ICM and the EM-like algorithms were both initialized with a simple $K$-means procedure. The stopping criterion is then settled to a number of 200 iterations that is enough to ensure the convergence of the procedure. 

In what follows, we restrict each $A(\ell)$ to be of the same dimension and in particular square block of dimension $b\times b$. For the sake of clarity the Block Likelihood Criterion is indexed by the dimension of the blocks, namely for a partition of square blocks of size $b\times b$ for which $\tilde{\x}=\tilde{\x}^{\mfl}$ and $\hat{\theta}_{\mle}=\hat{\theta}^{\mfl}$, we note it $\BLIC^{\mfl}_{b\times b}$. As already mentioned, we then have $\BIC^{\mfl}=\BLIC_{1\times 1}^{\mfl}$. We recall that when $B(\ell)=\emptyset$, $\tilde{\x}$ is omitted in the previous notations, that is for a square blocks partition we note our criterion $\BLIC_{b\times b}$. Then $\BLIC_{1\times 1}$ is the BIC approximations corresponding to a finite independent mixture model. All criterion were tested on simulated images obtained using the Swendsen-Wang algorithm.  We describe below the different experiments settings we have considered and the results we got.

\subsection{First experiment: selection of the number of colors}
\label{subsec:first-xp}

In this experiment the dependency structure is assumed to be known and the aim is to recover the number $K$ of colors of the latent configuration. We considered realizations from hidden Potts models with $K_T=4$ colors and $\sigma_k=0.5$. The interaction parameter $\psi$ was set close to the phase transition, namely $\psi = 1$ and $\psi=0.4$ for $\G_4$ and $\G_8$ respectively. These values of the parameter ensure the images present homogeneous regions and then the observations exhibit some spatial structure. Such settings illustrate the advantage of taking into account spatial information of the model. Obviously, for values of $\psi$ where the interaction is weaker, the benefit of the criterion that include the dependency structure of the model is not clear. The latter could even be misleading in comparison with BIC approximations for independent mixture models when $\psi$ is close to zero. On the other side, when $\psi$ is above the phase transition, the distribution on $\X$ becomes heavily multi-modal and there is almost solely one class represented in the image regardless the number of colors of the model. We carried out 100 simulations from the first order neighborhood structure $\G_4$ and 100 simulations from the second order neighborhood structure $\G_8$.

\begin{table}[H]
\caption{\small Selected $K$ in the first experiment for 100 realizations from ${\rm HPM}(\G_4,\theta, 4)$ and 100 realizations from ${\rm HPM}(\G_8,\theta, 4)$ using Pseudolikelihood Information Criterion ($\PLIC$), mean field-like approximations ($\BIC^{\mfl}$, $\BIC^{\gbf}$) and Block Likelihood Information Criterion ($\BLIC$) for various sizes of blocks and border conditions.}
\renewcommand{\arraystretch}{1.2}
\label{tab:select-K}
\centering
\begin{minipage}[t]{7.2cm}
\centering
\begin{tabular}{l|cccccc}
\multicolumn{1}{c}{}& \multicolumn{6}{c}{${\rm HPM}(\G_4,\theta, 4)$} \\
\hline
{\bf K} & {\bf 2} & {\bf 3} & {\bf \textcolor{red}{4}} & {\bf 5} & {\bf 6} & {\bf 7} \\
 \hline
 \hline
\rowcolor[gray]{0.85} $\PLIC$ & 0 & 9 & \textcolor{red}{91} & 0 & 0 & 0  \\
$\BIC^{\mfl}$ & 0 & 0 & \textcolor{red}{39} & 23 & 16 & 22 \\
\rowcolor[gray]{0.85} $\BIC^{\gbf}$ & 0 & 0 & \textcolor{red}{39} & 25 & 18 & 18 \\
$\BLIC^{\mfl}_{2\times 2}$ & 0 & 0 & \textcolor{red}{58} & 18 & 8 & 16\\
\rowcolor[gray]{0.85} $\BLIC_{1\times 1}$ & 0 & 0 & \textcolor{red}{97} & 1 & 2 & 0\\
$\BLIC_{2\times 2}$ & 0 & 0 & \textcolor{red}{100} & 0 & 0 & 0  \\
\multicolumn{1}{c}{}& \multicolumn{6}{c}{}
\end{tabular}
\end{minipage}
\begin{minipage}[t]{7.2cm}
\centering
\begin{tabular}{l|cccccc}
\multicolumn{1}{c}{}& \multicolumn{6}{c}{${\rm HPM}(\G_8,\theta, 4)$} \\
\hline
{\bf K} & {\bf 2} & {\bf 3} & {\bf \textcolor{red}{4}} & {\bf 5} & {\bf 6} & {\bf 7} \\
 \hline
 \hline
\rowcolor[gray]{0.85} $\PLIC$ & 0 & 7 & \textcolor{red}{93} & 0 & 0 & 0  \\
$\BIC^{\mfl}$ & 0 & 0 & \textcolor{red}{43} & 18 & 19 & 20 \\
\rowcolor[gray]{0.85} $\BIC^{\gbf}$ & 0 & 0 & \textcolor{red}{52} & 20 & 19 & 9 \\
$\BLIC^{\mfl}_{2\times 2}$ & 0 & 0 & \textcolor{red}{52} & 14 & 17 & 17\\
\rowcolor[gray]{0.85} $\BLIC_{1\times 1}$ & 0 & 3 & \textcolor{red}{90} & 1 & 4 & 2\\
$\BLIC_{2\times 2}$ & 0 & 1 & \textcolor{red}{99} & 0 & 0 & 0  \\
\rowcolor[gray]{0.85} $\BLIC_{4\times 4}$ & 0 & 0 & \textcolor{red}{100} & 0 & 0 & 0
\end{tabular}
\end{minipage}
\end{table}

\begin{figure}[H]
\centering
\begin{minipage}[t]{\textwidth}
\centering
\includegraphics[width=8cm] {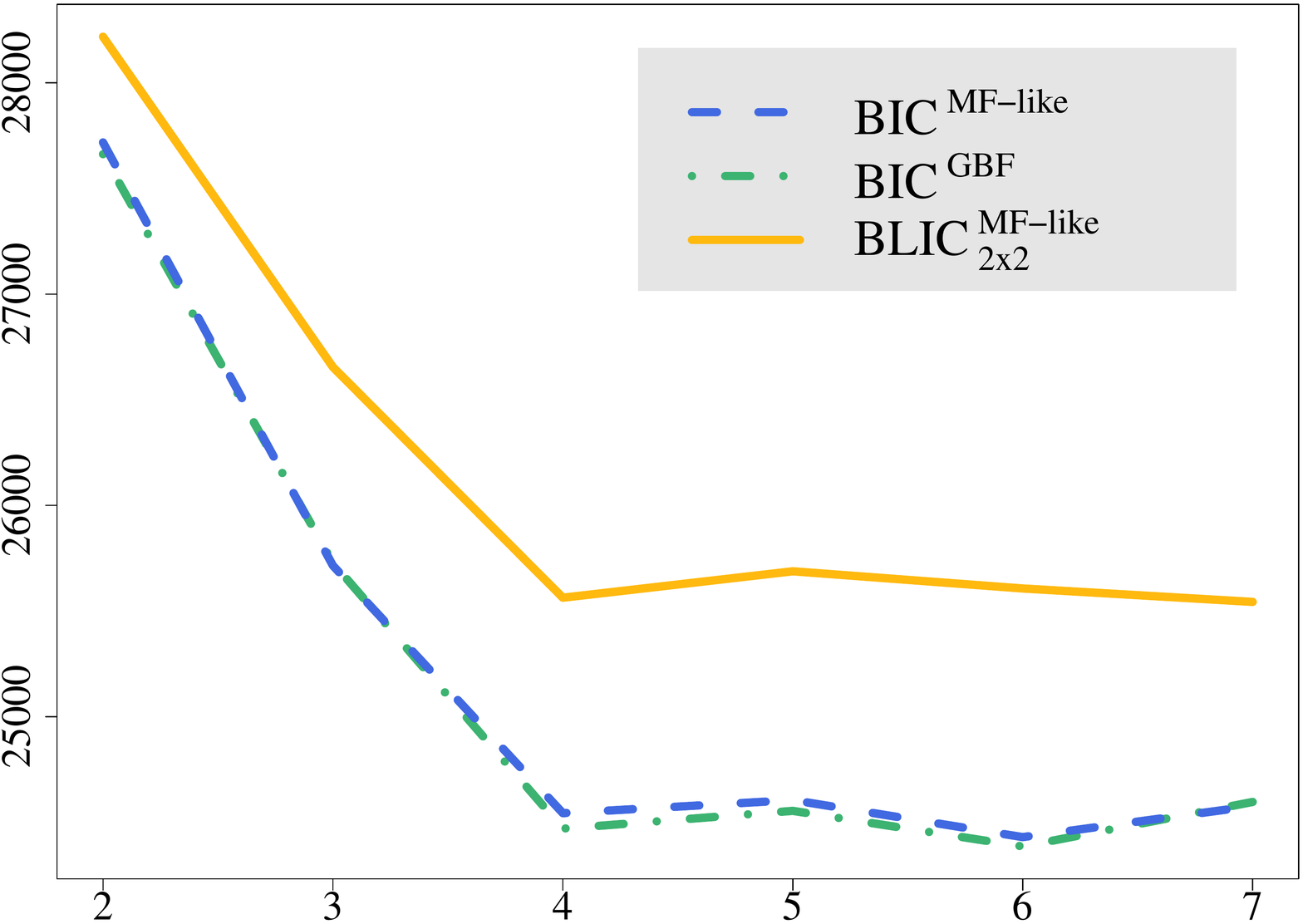}\\
(a)

\medskip
\end{minipage}
\\

\begin{minipage}[t]{7.3cm}
\centering
\includegraphics[width=7.3cm] {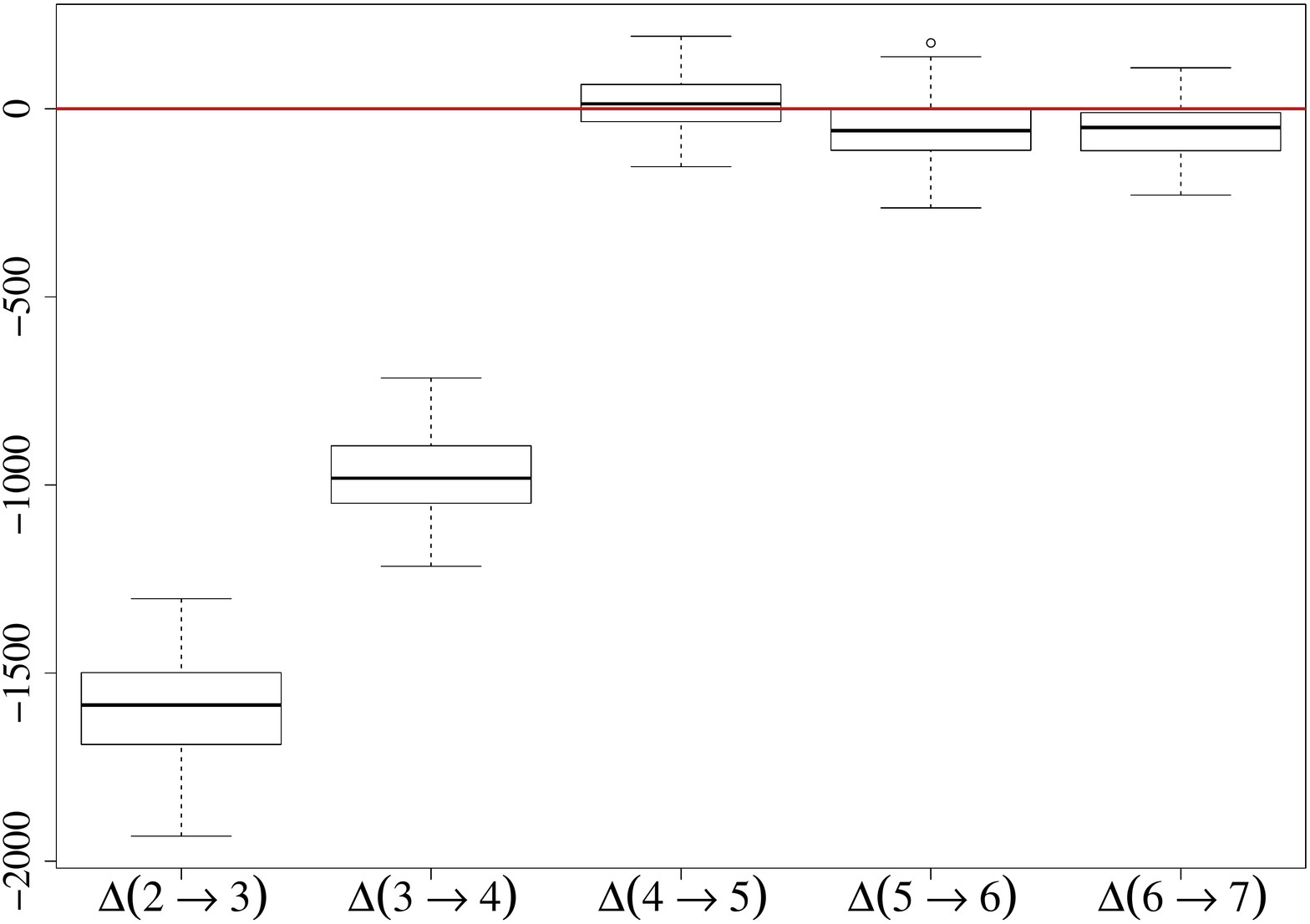}\\
(b)
\end{minipage}
\begin{minipage}[t]{7.3cm}
\centering
\includegraphics[width=7.3cm] {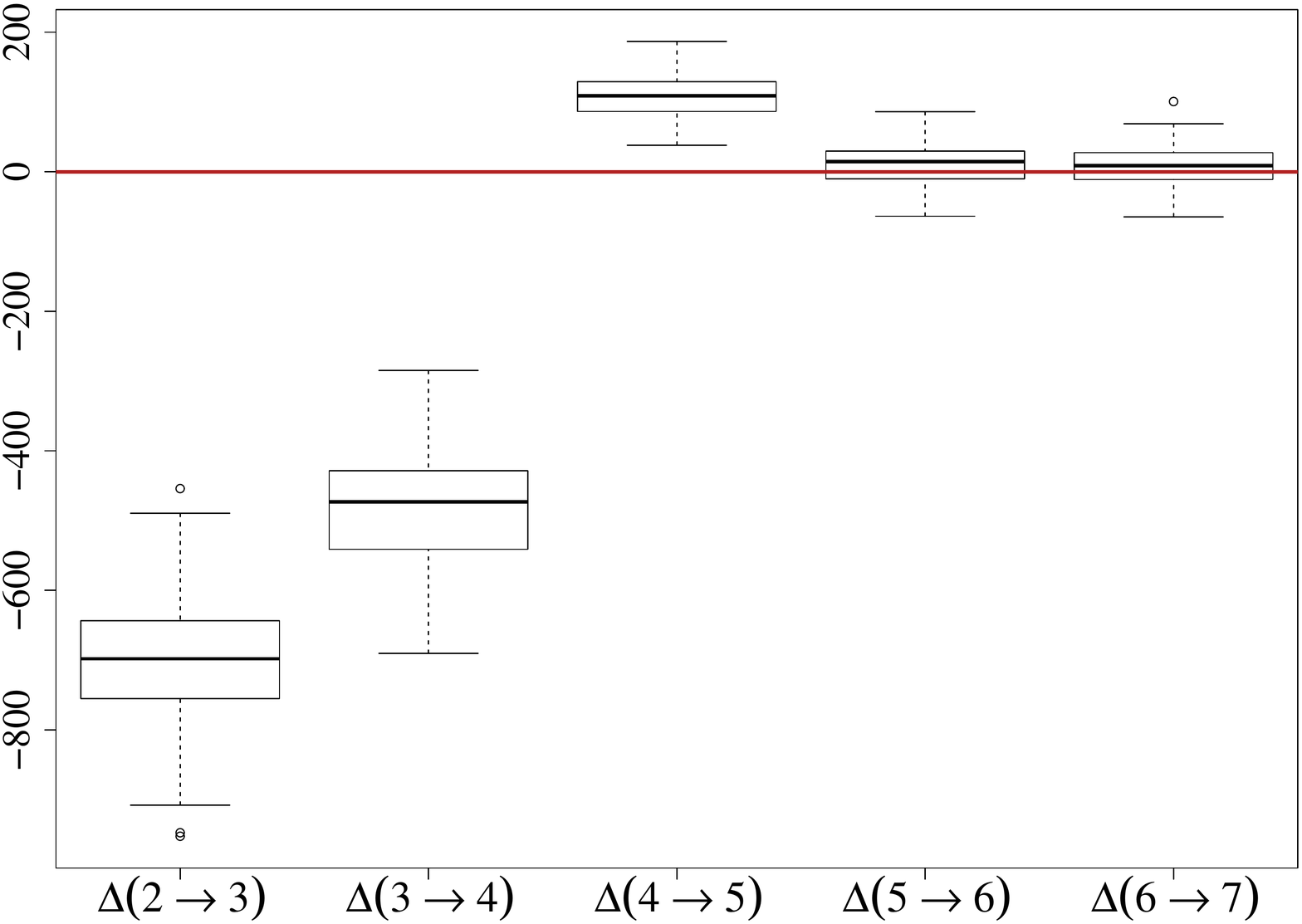}\\
(c)
\end{minipage}
\caption{\small \textbf{First experiment results.} (a) $\BIC^{\mfl}$, $\BIC^{\gbf}$ and $\BLIC_{2\times 2}^{\mfl}$ values for one realization of a first order hidden Potts model ${\rm HPM(\G,\theta,4)}$.
(b) Difference between $\BLIC^{\mfl}_{2\times 2}$ values for 100 realization of a first order hidden Potts model ${\rm HPM}(\G_4,\theta,4)$ as $K$ is increasing.
(c) Difference between $\BLIC_{2\times 2}$ values for 100 realization of a first order hidden Potts model ${\rm HPM}(\G_4,\theta,4)$ as $K$ is increasing
}
\label{fig:bic-plateau}
\end{figure}

The results obtained for the different criterion are reported in Table \ref{tab:select-K}. For $b\geq 2$, $\BLIC_{b\times b}$ outperform the different criterion even though $\PLIC$ and $\BLIC_{1\times 1}$ provide good results. By contrast approximations based  on mean field-like approximations, that is $\BIC^{\mfl}$, $\BIC^{\gbf}$ and $\BLIC_{2 \times 2}^{\mfl}$, perform poorly. These conclusions need nonetheless to be put into perspective. Figure \ref{fig:bic-plateau}(a) shows that the main issue encountered by these criterion is their inability to discriminate between the more complex models. Indeed these BIC approximations reach a plateau from $K=4$, a problem that other criterion do not face. As an example, Figure \ref{fig:bic-plateau}(b) and Figure \ref{fig:bic-plateau}(c) represent boxplots of the difference between BIC values for ${\rm HPM}(\G_4,\theta,K)$ as $K$ is increasing for the 100 realizations, namely
\[
\Delta\left(K\rightarrow K+1\right)=\BIC\left({\rm HPM}(\G,\hat{\theta}_{\mle},K+1)\right)-\BIC\left({\rm HPM}(\G,\hat{\theta}_{\mle},K)\right),
\]
for $K=K_{\rm min},\ldots,K_{\rm max}$. Hence, BIC approximations grow with $K$ if $\Delta\left(K\rightarrow K+1\right)\geq 0$ and decrease otherwise. It appears that $\BLIC_{2\times 2}$ increases systematically from $K=4$ whereas $\BIC^{\mfl}$ tend to be constant, or even decreases, so that none minimum can be clearly identified. We do not provide the boxplots for $\BIC^{\mfl}$ and $\BIC^{\gbf}$ because they are significantly the same. 

Finally these results illustrate in particular the importance of a well chosen segmented field $\tilde{\x}$. Indeed $\PLIC$ and $\BIC^{\mfl}$ are both criterion of type $\BLIC_{1\times 1}^{\tilde{\x}}$ but their performances greatly differ on this example. As regards the selection of $K$, $\BLIC_{b\times b}$ circumvent this question whilst performing better.

\subsection{Second experiment: selection of the dependency structure}
\label{subsec:second-xp}

For this second experiments the setting was exactly the same than for the first experiment. The only difference is that as first instance the number of colors $K_T$ is assumed to be known while the neighborhood system has to be chosen. To answer such a question it is obvious that we can not use criterion $\BLIC_{1\times 1}$ based on independent mixture model. 

As regards this question, all but two criterion perform very well, see Table \ref{tab:select-G}. In the first place, PLIC faces trouble to select the correct $\G_4$. This illustrate the importance of the estimation of the interaction parameter $\psi$. We have observed that the ICM algorithm whilst providing good segmented field, produces poorer estimates of the parameter than the simulated field algorithm. This has an impact quite important since $\psi$ sets the strength of interaction between neighboring nodes of the graph $\G$ and is most representative of the spatial correlation. On the other hand, $\BLIC_{2\times 2}$ fails to select the neighborhood system for second order hidden Potts model ${\rm HPM}(\G_8,\theta,4)$. This conclusion can be simply explained by the fact that the block does not include enough spatial information to discriminate between the competing models. When the primary purpose is the selection of a dependency structure, we should use block large enough to be informative regarding the different neighborhood systems in competition.

\begin{table}{H}
\caption{\small Selected $\G$ in the second experiment for 100 realizations from ${\rm HPM}(\G_4,\theta, 4)$ and 100 realizations from ${\rm HPM}(\G_8,\theta, 4)$ using Pseudolikelihood Information Criterion ($\PLIC$), mean field-like approximations ($\BIC^{\mfl}$, $\BIC^{\gbf}$) and Block Likelihood Information Criterion ($\BLIC$) for various sizes of blocks and border conditions.}
\renewcommand{\arraystretch}{1.2}
\label{tab:select-G}
\centering
\begin{minipage}[t]{7cm}
\centering
\begin{tabular}{l|cc}
\multicolumn{3}{c}{${\rm HPM}(\G_4,\theta, 4)$} \\
\hline
\multicolumn{1}{c}{} & \textcolor{red}{\bf $\G_4$} & {\bf $\G_8$} \\
\hline
\hline
\rowcolor[gray]{0.85} $\PLIC$ & \textcolor{red}{53} & 47 \\
$\BIC^{\mfl}$ & \textcolor{red}{100} & 0 \\
\rowcolor[gray]{0.85} $\BIC^{\gbf}$ & \textcolor{red}{100} & 0 \\
$\BLIC^{\mfl}_{2\times 2}$  & \textcolor{red}{100} & 0 \\
\rowcolor[gray]{0.85} $\BLIC_{2\times 2}$  & \textcolor{red}{100} & 0 \\
\multicolumn{1}{c}{}& \multicolumn{2}{c}{}
\end{tabular}
\end{minipage}
\begin{minipage}[t]{7cm}
\centering
\begin{tabular}{l|cc}
\multicolumn{3}{c}{${\rm HPM}(\G_8,\theta, 4)$} \\
\hline
\multicolumn{1}{c}{}  & {\bf $\G_4$} & \textcolor{red}{\bf $\G_8$} \\
 \hline
 \hline
\rowcolor[gray]{0.85} $\PLIC$ & 0 & \textcolor{red}{100} \\
$\BIC^{\mfl}$ & 0 & \textcolor{red}{100} \\
\rowcolor[gray]{0.85} $\BIC^{\gbf}$ & 0 & \textcolor{red}{100} \\
$\BLIC^{\mfl}_{2\times 2}$ & 0 & \textcolor{red}{100}\\
\rowcolor[gray]{0.85}  $\BLIC_{2\times 2}$  & 59 & \textcolor{red}{41} \\
$\BLIC_{4\times 4}$  & 0 & \textcolor{red}{100}
\end{tabular}
\end{minipage}
\end{table}

Aside the two above exceptions, the good performances of all criteria can be surprising. The same experiment has been done for stronger noise with $\sigma_k =0.75$ and $\sigma_k=1$. The conclusion remains the same. It appears that for a conditionally independent noise process, neighborhood system are readily distinguished close to the phase transition. This is not true for any parameter value as illustrated in the third experiment.

In the second instance, we supposed that $K_T$ and $\G$ were unknown, so that we were interested in the joint selection of the number of colors and of the dependency graph. For this example, the results remain the same than in Table \ref{tab:select-K} with the exception of PLIC. Indeed, the different criterion manage to differentiate the model in terms of the graph $\G$ so that their performances are directly related to their ability to choose the correct number of colors. 

\subsection{Third experiment: BLIC \textit{versus} ABC}
\label{subsec:third-xp}

This third experiment is the occasion to compare BLIC with the ABC procedures proposed by \cite{stoehr2015}. We return to the problem of solely selecting the dependency graph when the number of colors is known. We still consider a homoscedastic Gaussian noise but over bicolor Potts models (K=2). The standard deviation $\sigma_k=0.39$, $k\in\{0,1\}$, was set so that the probability of a wrong prediction of the latent color with a marginal MAP rule on the Gaussian model is about $10\%$ in the thresholding step of the ABC procedure. Regarding the dependency parameter $\psi$, we set prior distributions below the phase transition which occurs at different levels depending on the neighborhood structure. Precisely we used  a uniform distribution over $(0;1)$ when the adjacency is given by $\G_4$ and a uniform distribution over $(0; 0.35)$ with $\G_8$.
In order to examine the performance of model choice criteria in comparison of ABC, we carried out 1000 realizations from ${\rm HPM}(\G_4, \theta, 2)$ and 1000 realizations from ${\rm HPM}(\G_8, \theta, 2)$ with parameters from the priors. The results are presented in Table \ref{tab:comparaison}

\begin{table}{H}
\caption{\small Evaluation of the prior error rate of ABC procedures and of the error rate for the model choice criterion in the third experiment.}
\label{tab:comparaison}
\centering
\renewcommand{\arraystretch}{1.2}
 \begin{tabular}{lcc||lc}
    \hline
  \textbf{Train size} & $\bf 5,000$ & $\bf 100,000$ & \textbf{Criterion} & \textbf{Error rate}
    \\
    \hline
    \hline
     \rowcolor[gray]{0.85} 2D statistics & $14.2 \%$  & $13.8 \%$ & $\PLIC$ & $19.8\%$
    \\
    4D statistics & $10.8 \%$ & $9.8 \%$ & $\BIC^{\mfl}$ & $7.6\%$
    \\
  \rowcolor[gray]{0.85}   6D statistics & $8.6 \%$ & \textcolor{red}{$6.9 \%$}  & $\BIC^{\gbf}$ & $7.1\%$
    \\
    Adaptive ABC & $8.2 \%$ & \textcolor{red}{$6.7 \%$} &  $\BLIC_{4 \times 4}$ & $7.7\%$

\end{tabular}
\end{table}

The novel ABC procedure introduced by \cite{stoehr2015} appears to provide the best performances but for a training reference table of size 100 000. This reinforces the idea that for unlimited computation possibilities, ABC can efficiently address situations where the likelihood is intractable. However, Table \ref{tab:comparaison} suggest that for a much lower computational cost it is possible to get equivalent, or even better, error rate by using model choice criterion $\BIC^{\mfl}$, $\BIC^{\gbf}$ or $\BLIC_{b\times b}$, while $\PLIC$ seems not to be overtaken. In this example, $\BIC^{\gbf}$ slightly supersede $\BIC^{\mfl}$ and $\BLIC_{b\times b}$. This can be explained by the fact that for parameter from the prior close to zero, the assumption of independence between the sites is almost true. In the latter case, estimating BIC using the first order approximations of the partition function of Gibbs distribution \citep{forbes2003} may be preferable than using normalizing constants defined on blocks.

\section{Conclusion and perspective}

In the present article, we considered BIC to perform model selection when dealing with hidden Markov random fields. To avoid time consuming simulation methods like MCMC or ABC algorithms, we proposed to move towards variational methods and in particular to use valid probability distributions over non-overlapping blocks of the lattice in place of the intractable likelihood (Section \ref{subsec:gibbs-approx}). Consequently, we derived Block Likelihood Information Criterion to discriminate between hidden Markov random fields.

The numerical results (Section \ref{sec:blic-results}) highlighted that the approximations of BIC based on independent blocks without fixed border provide better performances comparing to pre-existing criteria as regards the inference of the number of latent colors. This conclusion has to be brought into perspective for the selection of the dependency structure as the size of the blocks should be wide enough unless BLIC can be misleading. According to the numerical results (Section \ref{subsec:third-xp}), the opportunity we have explored appears to be a satisfactory alternative to ABC model choice algorithms which besides their computational cost are delicate to calibrate \citep[\eg,][]{stoehr2015}. Our approach offers thus an appealing trade-off between efficient computation and reliable results.

While the numerical part of the paper assess its efficiency, the novel criterion makes in its current version two major approximations that are worth exploring. First mention, the choice of a particular substitute is lead by any optimality conditions. From that viewpoint, the construction of an optimal approximations regarding the variational free energy over the set of probability distributions that factorize on blocks is yet to be studied. The second level of approximations concerns the penalty term. The next step of our work cannot be reduced to the sole aim of improving the quality of the approximations. Through Section \ref{subsec:first-xp}, we have seen that an optimal solution with respect to the Kullback- Leibler divergence is not sufficient to ensure a good behaviour of model choice criteria, especially if the more complex model are not enough penalized. The penalty term used is solely valid for independent variable. We have neglected the interaction within a block, an assumption that slightly modified the number of free parameter. The impact of dependence variables on the penalty term is a logical follow-up to our work.


\end{document}